# Controllable chiral light generation and vortex field investigation using plasmonic holes revealed by cathodoluminescence

*Takumi Sannomiya[1]\*, Taeko Matsukata[1], Naoki Yamamoto[1]*

AUTHOR ADDRESS

[1] Department of Materials Science and Technology, Tokyo Institute of Technology, 4259 Nagatsuta Midoriku, Yokohama 226-8503, Japan

\*sannomiya.t.aa@m.titech.ac.jp

ABSTRACT

Control of the angular momentum of light is a key technology for next-generation nano-optical devices and optical communications, including quantum communication and encoding. We propose an approach to controllably generate circularly polarized light from a circular hole in a metal film using an electron beam by coherently exciting transition radiation and light scattering from the hole through surface plasmon polaritons. The circularly polarized light generation is confirmed by fully polarimetric four-dimensional cathodoluminescence mapping where angle-resolved spectra are simultaneously obtained. The obtained intensity and Stokes maps show clear interference fringes as well as almost fully circularly polarized light generation with controllable parities by electron beam position. By applying this approach to a metal film with three holes, a vortex field with a phase singularity is visualized in the middle of the holes.

# Main

  Control of angular momentum of light is one of the crucial technologies to realize next-generation nano-optical devices and optical communications, including quantum communication and encoding,[1-4] and has been extensively investigated in topological photonic structures or for coupling to the electron spins in materials. Spin angular momentum (SAM) of light is represented by left- or right-handed circularly polarized light (CPL), which is a robust signal carrier and has already been practically utilized e.g. in 3D cinemas to send two distinct signals for right and left eyes regardless of the observation angle. For a macroscopic CPL parity control, a linear polarizer and a quarter waveplate are typically combined as a set of filters, which requires much larger scale elements than the light wavelength. For micro or nanoscale devices, it has been proposed to couple a chiral optical nanoantenna with a light source to select the CPL parity by the structure.[5-7] This antenna coupling strategy is useful because a linearly polarized light source can be utilized, instead of a CPL-selective source which is not readily available, when coupled with such chiral antennas. However, when the light source is fixed to the structure, dynamic control of CPL is not straightforward because the chirality is fixed to the antenna geometry. To overcome this problem, achiral nanoantennas combined with an optical waveguide have been proposed, where the optical signal from different directions along the waveguide determines the generated CPL parity.[8] This is based on an elaborate fabrication of optical waveguides with coupled nanoantennas and highly efficient propagation of light is required. As an alternative approach, CPL generation using an electron beam, or cathodoluminescence (CL), has been proposed, which is possible even from a spherical particle by controlling the phase difference of two dipoles in the structure.[9-11] Such electron beam-based light generation is expected to be a scheme for prospective laser or quantum light sources.[12, 13] Although the use of the electron beam for optical devices requires different technologies compared to purely light-based systems and may give limitations, the electron-beam approach is not unrealistic as there have been various table-top devices with electron beams, such as cathode-ray tubes, streak-cameras, short-wavelength light sources, indicating the practical applicability of this approach. While integrated photonics facilitate the production of the light sources and would realize photonic circuits [7, 8, 14], electron-beam-based technology can offer extremely fast switching,[12] as is already used in fast cameras.

Here we propose a new approach to generate CPL using circular metallic holes coupled to surface plasmon polaritons (SPPs), that is a different mechanism compared to the previous studies. Plasmonic hole structures are useful antennas since SPP is already incorporated in the structure working as a guided wave on the surface. In such structures, transition radiation (TR) and SPP scattering field, which are generated at spatially separate locations, leads to interference; When an electron beam is irradiated onto the metal film, away from the antenna structures, TR is generated at the electron beam position and simultaneously and coherently excited SPPs are scattered by the antenna and creates photons. Since all the interaction processes are coherent with respect to the electron excitation, the light from these two sources interferes.[15] Such interferences with TR have given a way to access the phase of the field in the CL measurement.[16] In this study, we utilize this interference of TR and the light through SPP scattering at the hole to control CPL (Fig. 1a). Also using this CPL generation tool with an electron beam, we further visualize a vortex field generation from three holes, emulating a three-phase optical field motor, which corresponds to the orbital angular momentum (OAM) in the SPP field.

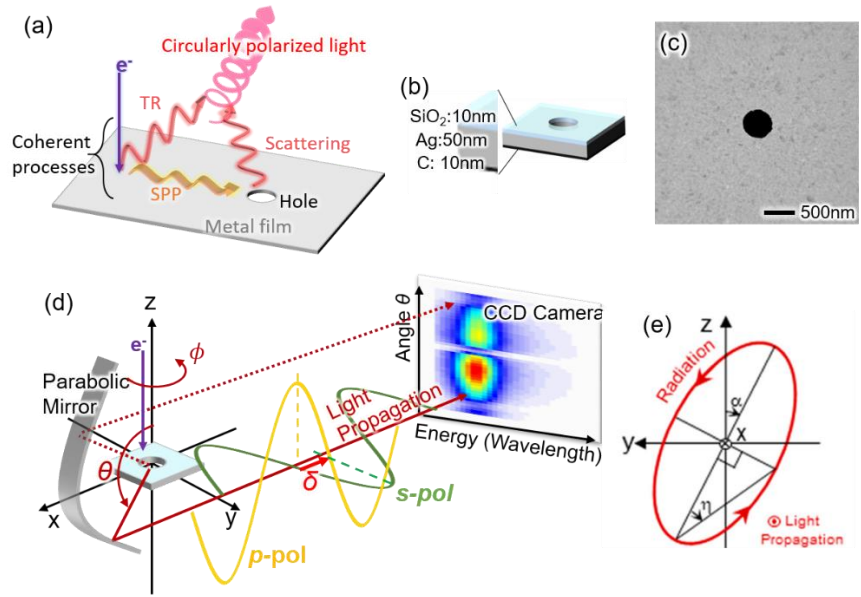

**Figure 1.** Concept and method of the study. (a) Illustration of circularly polarized light generation through interference of transition radiation (TR) and surface plasmon polariton (SPP) scattering from a hole excited by electron beam. (b) Dimension of the free-standing nanohole membrane used in the study. (c) STEM dark field image of a fabricated 500 nm hole. (d) Illustration of the fully polarimetric four-dimensional cathodoluminescence. The coordinate, angle, and polarization definitions are also shown. (e) Definition of polarization parameters.

Holes are fabricated on a free-standing silver membrane by colloidal lithography with a film transfer method.[17, 18] The "upside" of the silver membrane is covered by $SiO_2$ to avoid degradation, and the "downside" is covered by carbon so that SPP propagates only along the upside of the membrane (Fig. 1b). A scanning transmission electron microscopy (STEM) CL system is used to evaluate the CPL generation from circular silver holes. The four-dimensional (4D) Stokes data acquisition in this setup allows simultaneous angle- and wavelength-resolved CL mapping, as illustrated in Fig. 1d and e.[9]

Figure 2 shows CL mapping results of a 500 nm silver hole with *p*-polarization at various detection angles $\theta$ and photon wavelengths. These CL maps are obtained by scanning the electron beam two-dimensionally on the sample and the CL signal is plotted in synchronization with the beam scan. The left and right halves of the CL maps with symmetric detection angles are shown, namely pairs of $\theta = 45°$, $135°$ and pairs of $\theta = 30°$, $150°$, facing each other in order to compare their interference fringe patterns. The interference fringe pattern is formed clearly at the top and bottom of the images with different fridge pitches, which are also dependent on the wavelength and detection angles. These fringe patterns can be understood as the interference between TR and the scattering from the hole through SPP propagation, as schematically illustrated in Fig. 2a.[16] The phase difference $\varphi$ of TR and the scattered wave from the hole generates the interference fringe patterns. The phase difference $\varphi$ can be expressed using the electron beam position $R$ from the center of the hole as,

$$\varphi = R\left(k_{spp} + k\sin\theta\cos\phi\right) + \Delta \qquad (1)$$

where $k_{spp}$ is the SPP wave vector, $k$ is the free-space light wave vector, and $\phi$ is the azimuthal angle of the position. $\Delta$ corresponds to an additional phase shift related to the excitation of the TR and the scattering of the SPP by the hole. The interference patterns have shorter pitches in the upper half of the image ($x > 0$) than the lower half ($x < 0$), which is because the sign of the TR phase shift (second term in Eq.1) flips with respect to SPP-scattering light (first term in Eq.1), as illustrated in Fig. 2a. The comparison of the maps with symmetric detection angles with respect to the sample plane reveals that the fringe pattern does not match for the symmetric pairs ($\theta = 45°$, $135°$ pair in Fig. 2b and $\theta = 30°$, $150°$ pair in Fig. 2c). This mismatch is related to the phase difference of the TR towards the top ($\theta < 90°$) and bottom ($\theta > 90°$) spaces, contributing to the offset $\Delta$ in Eq.1. This phase difference between the top-ward and the bottom-ward TRs is approximately $\pi$ when the space is separated by an ideal metal film producing mirror-symmetric dipole radiations.

To better understand the behavior of TR, we calculated the electromagnetic field distribution by 80 keV electron moving in and out of a silver surface, as shown in Figure 2d.[19] One can see TR emitted into the free space and SPP propagating on the surface, as described in the scheme of Fig.2a. Since the TR is an effective electric dipole polarized perpendicular to the interface, the transverse magnetic field, as plotted in Fig. 2d, well represents the phase of the TR field. The zero phase is set when the electron is located at the interface. As clearly seen, the phase difference is ~$\pi$, which can be slightly shifted due to the presence of the carbon layer. The theoretically calculated TR phase from a carbon surface is also shown in the Supporting Information. While TR can be approximated by a perpendicular electric dipole, the field from a hole in a metallic film can be described by a magnetic dipole[20-23], which e.g. has the magnetic polarization along the *y*-axis when the electron beam is placed on the *x*-axis as in the configuration of Fig. 2a. This scattering field from the hole magnetic dipole gives symmetric magnetic distributions on the top (*z*-positive) and back (*z*-negative) sides of the sample, as shown on the right of Fig. 2d. Thus, the radiations by TR and hole scattering differently interfere in the top (*z*-positive) and back (*z*-negative) side hemispheres of the space, generating shifted interference patterns as in the experiment in Fig. 2b and c. In the upside-down configuration of the sample, as shown in Figure 2e, the interference pattern shifts only slightly. This indicates that the phase shifts of the SPP and TR excitations by flipping the sample cancels out the total phase offset delta together with the influence of the carbon layer.

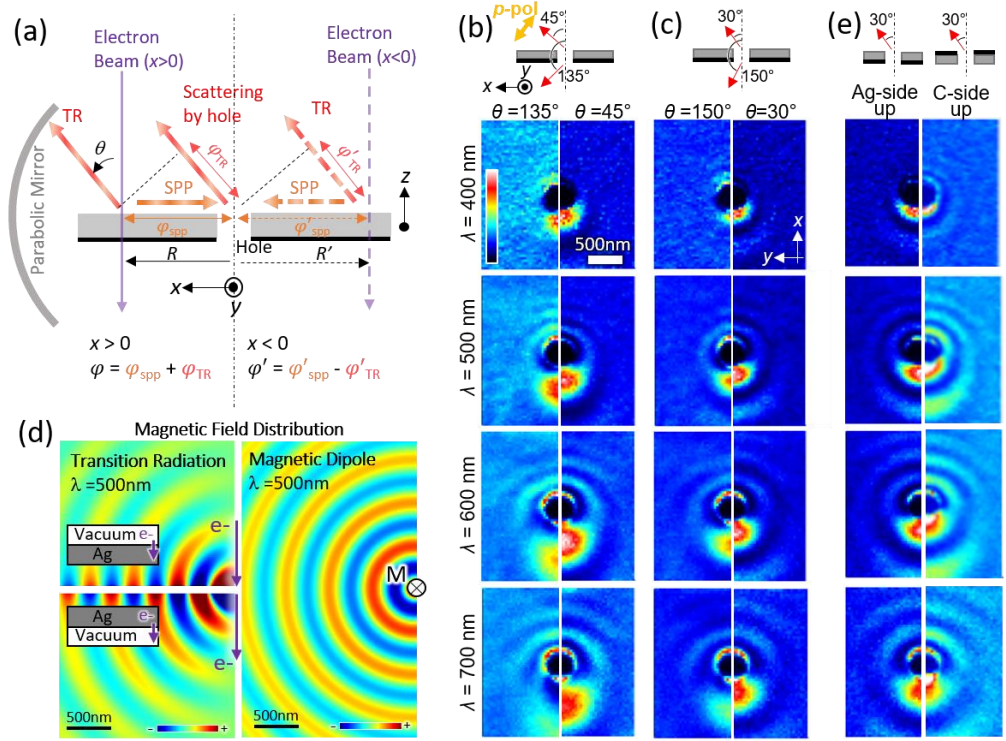

**Figure 2.** CL mapping of a 500 nm single silver hole with p-polarization, i.e. polarized in the *x-z* plane) (a) Schematic illustration of the interference between TR and scattering by the hole through SPPs projected on the *x-z* plane. The condition where the electron beam is incident at negative *x* positions is indicated by dashed lines and primes of the parameters. (b-c) Measurement results with halves of the images of symmetric detection pairs (b: $\theta$ =45° and 135°, c: $\theta$ =30° and 150°) facing each other so that the fringe patterns can be directly compared. (d) Calculated magnetic field perpendicular to the screen for (left) transition radiation and for (right) a magnetic dipole polarized perpendicular to the screen for the vacuum wavelength of 500 nm. TR is plotted for the upper (*z*-positive) and back (*z*-negative) sides of a bulk silver surface by 80 keV electron impact. (e) $\theta$ =30° measurement maps with the silver (left) and carbon (right) sides up. The mapping (b,c,e) are obtained by scanning the focused electron beam on the *x-y* plane on the sample and the image intensities are integrated over ± 25 nm in the wavelength and ± 2.5° in the angle $\theta$.

Considering that the interference between TR and SPP-mediated scattering gives nice fridge patterns for p-polarization mapping revealing the phase difference between TR and hole scattering, one can utilize this interference to produce CPL when the s-polarization component in the hole scattering is included. Although CPL cannot be produced in a system with a point symmetry, the symmetry of this circular nanohole system can be broken by the electron beam position, detection angle, and polarization, i.e. by so-called extrinsic chirality.[9, 10] We analyze these CPL properties by calculating Stokes parameters from six CL mapping images with different polarizations, namely p-polarization, s-polarization, +45°-polarization, -45°-polarization, RCP, and LCP (see Fig. 1d, e for the parameter definition). The six images with different polarizations are obtained separately by scanning the electron beam over the same area of the sample and the exact position of each image is aligned and normalized for the Stokes parameter calculation. More details of the procedure is described in Supporting Information as well as in the previous study.[9] The representative Stokes mapping results are shown in Fig. 3 for different detection angles and wavelengths. In all the maps, interference fringe patterns due to TR and hole scattering are visible. Similarly to the interference patterns in Fig. 2, these fringe patterns become asymmetric with respect to the y-axis (horizontal axis) when the detection angle is off the z-axis, as suggested by eq. 1.

In the S3 (RCP-LCP signal) maps in Fig. 3, both LCP (blue) and RCP (red) generations are confirmed, which are all mirror-symmetric with a flipped sign with respect to the *x-z* plane (vertical line at the center of the image). This symmetry can be understood from the inclusion of the mirror-symmetric electric field with a flipped sign against the *x-z* plane of the hole magnetic dipole polarized along the *x* direction when the electron beam is located off the center of the hole. Since the electric field of TR is symmetric with respect to the *x-z* plane, the interference between the TR and hole scattering generates CPL together with the interference fringe patterns corresponding to the phase difference between TR and hole scattering.

The $\delta$ maps, showing the phase difference between the p-polarization and s-polarization components, and the $\eta$ maps, corresponding to CPL "circularity", show that almost perfect CPL ($\delta = \pm \pi/2$ with $\eta = \pm \pi/4$) is available.(see Fig. 1d,e for the parameter definition) Under certain conditions especially with the electron beam situated at the bottom half of the image or near the hole, fully circular CPL can be observed. This shows that the polarization parameters of the generated CPL can be controlled by electron beam position for certain detection angles. Thanks to the SPP decay in the 2D planner space, the CPL intensity can be also controlled by the electron beam position by locating the beam closer or further to the hole. (see Supporting Information for the intensity discussion)

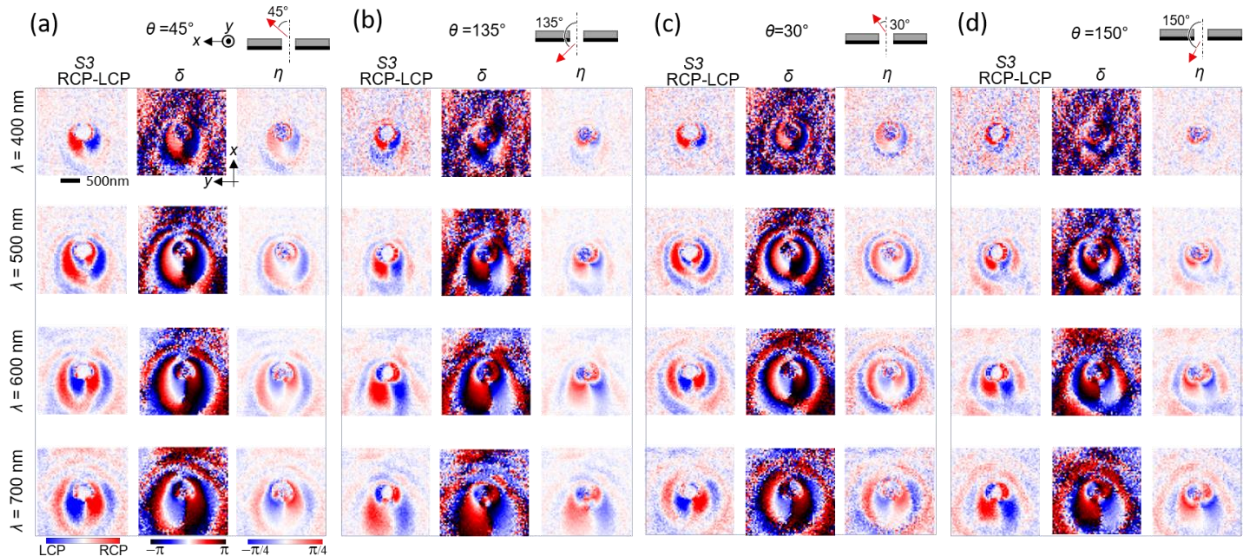

**Figure 3.** Circularly polarized light generation from a 500 nm single silver hole in various conditions. Six CL maps with different polarizations are obtained by scanning the focused electron

beam on the *x-y* plane of the sample and Stokes parameters are calculated at each pixel after adjusting the images positions. Stokes maps of *S3* (difference of RCP and LCP map intensities), $\delta$ and $\eta$ are performed for the detection angles of (a) $\theta = 45°$, (b) $\theta = 135°$, (c) $\theta = 30°$, and (d) $\theta = 150°$. Maps of representative photon wavelengths are listed in the rows.

While the radiation field from a hole in a metal film corresponds to the field distribution of a magnetic dipole,[20-23] the SPP field on the film generated by the hole can be treated as an in-plane-polarized electric dipole in the 2D space. We here utilize this hole dipole on the surface as an SPP field source to generate a vortex field on a surface, corresponding OAM in the SPP field, and visualize this field rotation by Stokes mapping. We demonstrate this vortex generation using three holes aligned as shown in Fig. 4. We chose this configuration with one large hole and two smaller ones to ascertain a strong SPP scattering field in the middle of the three with different phases to create a vortex. When a plane wave is illuminated, such holes within the illumination field will be coherently excited with some phase differences for a given illumination angle. Due to the reciprocity of the wave, CL measurement with a certain detection angle $\theta$ corresponds to the *z*-field measurement generated by such plane wave illumination with an angle $\theta$, as illustrated in Fig.4a. (see also the Supporting Information) Thus, we can reasonably analyze the wave interference from the three holes by CL field mapping as the generated field from the holes reciprocally. Such a rotating field from three field sources with certain phase differences can be understood from the concept of a three-phase motor, as described in Fig. 4b.

The measured CL maps with $\theta = 30°$ and $\lambda = 660$ nm are shown in Fig. 4c-g for different polarization conditions and representative Stokes parameters. The Stokes maps are reproduced from six images of different polarizations by scanning the electron beam over the sample, in the same manner as the results of Fig.3. Clearly, interferences of SPPs from the holes are observed in the positions between the holes, generating a more complicated field distribution compared to the single hole case. Closely looking at the s-pol field map and the $\delta$ map (Fig. 4g), one finds a position with a phase singularity in the middle of the three holes, which is marked by dashed circles. At this position, we expect field interferences of these three sources. The phase pattern (Fig. 4g) gives a rotating phase and no amplitude (Fig. 4d) at the center of this rotation is found, which are the typical features of the phase singularity of the vortex fields.

To compare with the experimental results and confirm the vortex field, analytically calculated field distributions are shown in Fig. 4h-m. In this model, three dipole sources on a 2D plane are excited by a plane wave with an incidence angle of $\theta = 30°$, which is reciprocal to the CL measurement (Fig. 4a). To introduce the additional field of the TR in the analytical calculation, a plane wave field over the surface of the calculation plane is overlaid (see calculation details in Supporting Information). The analytical calculation patterns in Fig. 4j-m well reproduce the experimental results, as seen in the pitch and position of the interference fringes in the p-pol and S3 maps (Fig. 4d, e, j, k) as well as the field singularity in the s-pol maps (Fig. 4d, j) at the vortex position indicated by the circle and the vortex phase rotation in the $\delta$ phase maps (Fig. 4g, m). Although only the phase difference of s- and p-polarizations (not each phase) is available in the experiment, according to the analytical calculation, the p-pol phase in Fig. 4i is dominated by TR (plane wave on the surface) constantly shifting in the *x* direction, with almost no local variation originating from the p-polarized dipole field from each hole. Thus, this CPL phase measurement basically extracts the phase of s-polarized (horizontally polarized) dipoles. The local variation of the relative phase δ (Fig. 4g, m) corresponds to the s-polarization phase (Fig. 4h) spread on the homogeneous background with a constant inclination of the p-polarized TR phase. (Fig. 4i).

At the vortex field position, as marked by circles in Fig. 4j, m in the middle of the three holes, the simulated singularity features well correspond to the experimental ones in the experiment (Fig. 4 d,g). Since this field singularity is mainly of the s-polarized (horizontally polarized) dipole fields (Fig.4h), it is natural that only the s-pol field gives zero amplitude at this singularity point (Fig. 4d,j) while the field in non-zero at this position in the p-pol field patterns (Fig. 4e,k). This singularity at the vortex center is also confirmed by the well-corresponding phase plots of s-pol and δ in Fig. 4h, m. In SI, the rotating phase reproduced by simulation is shown. Thus, we have visualized the SPP vortex field generated from three holes, demonstrating a three-phase field motor. This demonstrates that such a vortex field can be generated in a much simpler way than highly organized nanostructures.[24]

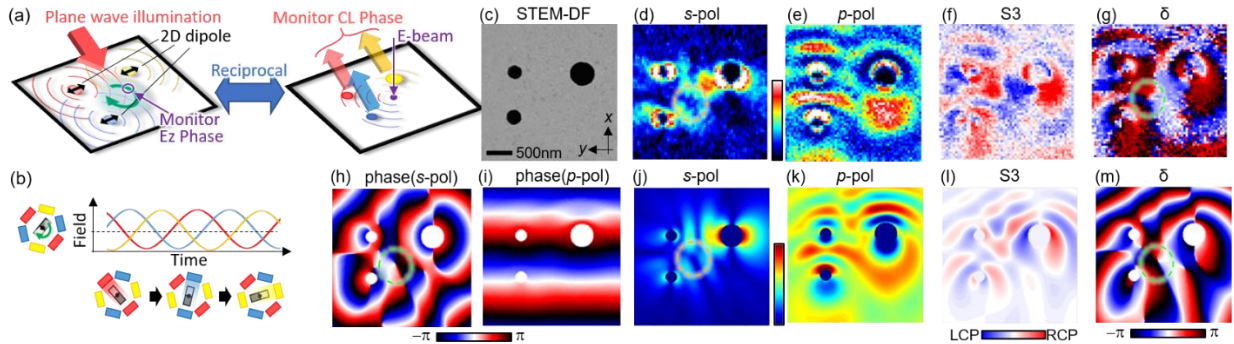

**Figure 4.** Vortex field generation using three holes. (a) Schematic illustration of monitoring the SPP field rotation generated by three holes with plane wave excitation and corresponding CL detection, which are reciprocal to each other. (b) Illustration of the field rotation from three sources, mimicking a three-phase motor. (c-g) Experimental Stokes maps obtained at the detection angle $\theta$

= 30° with the wavelength of 660 nm by scanning the electron beam on the *x-y* plane. The phase differences are generated by the *x* position difference dependent on *θ* and by the size difference of the hole. (h-m) Calculated field patterns with three in-plane electric dipoles and a propagating plane wave along the x-axis in a two-dimensional system, emulating the three holes and the TR in the experiment, respectively. The vortex position is indicated by dashed circles, in the *s*-polarization (d,j) and phase maps (g,h,m), corresponding to the *s*-polarization field rotation. The vortex center corresponds to a field singularity and has no field amplitude in the *s*-polarization map (d, j).

In conclusion, we have successfully demonstrated CPL generation from a hole in a metal film using an electron beam. This CPL is tunable by controlling the interference of TR at the electron beam position and coherent scattering field at the hole through SPP. The relative phase difference of the TR and hole scattering field determines the CPL parity. The CPL generation is confirmed by 4D Stokes mapping CL approach by visualizing the Stokes parameter distributions as functions of electron beam position, radiation angle, and wavelength. Using this CPL generation and mapping technique, a three-phase SPP field motor is also demonstrated where three dipole fields interfere and generate a rotating field, showing phase singularity. The CPL generation mechanism and field control strategy using coherent light generation processes are useful to realize on-demand CPL sources. Compared to the CPL generation by controlling the electron beam within nanoantennas of subwavelength scales,[9] the proposed approach requires less precise control of the electron beam size and position, i.e. in the scale of the light wavelength or even larger.

## Methods

For the CL measurement, a modified STEM (2100F, JEOL Japan) instrument equipped with a Schottky type field emission gun and an aberration corrector is operated at 80kV acceleration with a beam current of 1 nA.[9, 25] The parabolic mirror situated at the sample position collimates the light emission from the sample. Only the $\phi = 0°$ emission angle component of the emission is selected by the slit mask while $\theta = 0 - 180°$ component is dispersed on the CCD camera to obtain 2D information of $\theta$ and the wavelength at each electron beam position.(Fig. 1) With the electron beam scan, 4D data sets are obtained. A polarizer and a phase plate are inserted in between for the polarimetric analysis (Supporting Information). The dispersion of the phase plate and the phase shift due to reflection by the mirror are corrected.

## Acknowledgements

This works was financially supported by JSPS PD2(20J14821), JSPS Kakenhi (21H01782, 22H05033), and JST FOREST (JPMJFR213J). The authors thank Mr. S. Ogura for helping us with the CPL analysis.

## Corresponding author

Takumi Sannomiya : sannomiya.t.aa@m.titech.ac.jp

# Supporting information

Calculated transition radiation fields, analytical model based on reciprocity of CL detection and plane wave illumination, calculated phase rotation at singularity position, CL spectrum of a hole, Stokes mapping procedures, and signal intensity are available.

**TOC graphic:**

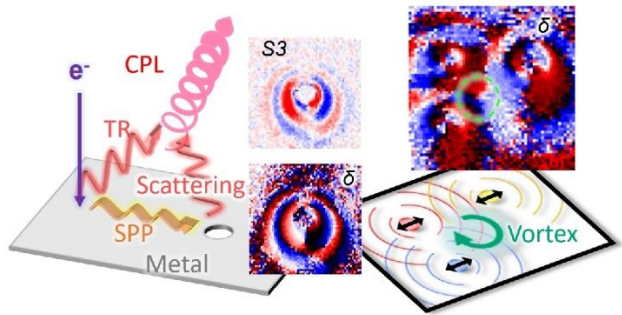

# Supporting Information for

Controllable chiral light generation and vortex field investigation using plasmonic holes revealed by cathodoluminescence

*Takumi Sannomiya[1], Taeko Matsukata[1], Naoki Yamamoto[1]*

AUTHOR ADDRESS

[1] Department of Materials Science and Technology, Tokyo Institute of Technology, 4259 Nagatsuta Midoriku, Yokohama 226-8503, Japan

## S1. Calculated transition radiation field

In Fig.2 in the main text, calculated transition radiation fields (TR) for a silver surface at the photon wavelength of 500 nm are shown. Here, we show TR fields in different conditions such as for a bottom carbon surface (Fig.S1a) and for a different photon wavelength (Fig.S1b). TR from carbon (Fig.S1a, left) shows a shifted phase compared to TR from silver (Fig.2d). Thus, inclusion of carbon as the surface layer would shift the TR at the bottom from the bulk silver. We also note that the propagation of the surface plasmon polariton (SPP) on the carbon surface is suppressed due to the absorptive nature of carbon.

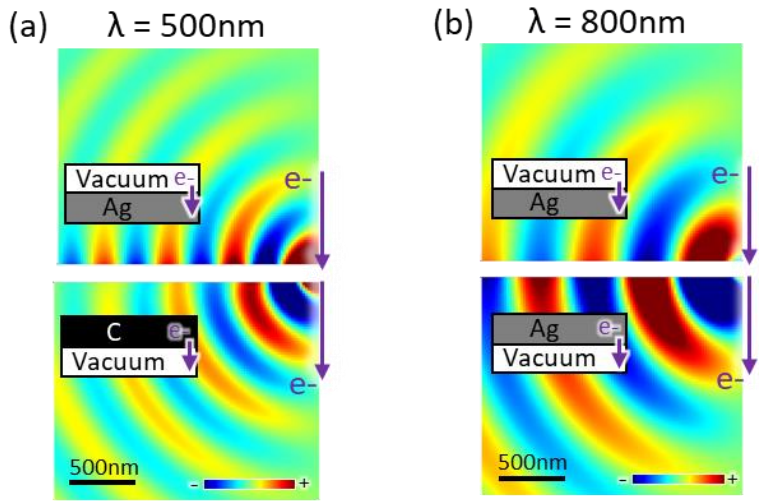

**Figure S1.** Calculated magnetic fields of transition radiation perpendicular to the screen. (a) Upper radiation from a bulk silver surface and lower radiation from a bulk carbon surface, calculated for the wavelength of 500 nm. (b) Upper and lower radiations from a bulk silver surface calculated for the wavelength of 800 nm.

## S2. Analytical model based on reciprocity of CL detection and plane wave illumination

Considering the reciprocity between cathodoluminescence (CL) detection and plane wave illumination, as described in Fig.4a in the main text and Fig.S2a below, we model the three-hole system of Fig.4 in the main text by three dipole sources representing the holes and a plane wave representing TR at each film position (Fig.2Sb). The field $E$ at position $\mathbf{r}$ on the film surface can be represented as,

$$E(\mathbf{r}) = \sum_{n=1}^{3} A_n \frac{\exp[i\{\mathbf{k}_{\mathrm{spp}} \cdot (\mathbf{r} - \mathbf{r}_n) + \mathbf{k} \cdot \mathbf{r}_n\}]}{\sqrt{r}} + A_{\mathrm{TR}} \exp(i\mathbf{k} \cdot \mathbf{r}).$$

The first sum term corresponds to the three 2D dipoles located at position $\mathbf{r}_n$ and the second term to the plane wave (or transition radiation). $\mathbf{k}_{\mathrm{spp}}$ and $\mathbf{k}$ are the wavevectors of the surface plasmon polariton and of the plane wave respectively. The coefficient $A$ is the complex amplitude of each wave source. For 2D dipoles, this coefficient includes the in-plane azimuthal angular dependence according to the polarization. To reproduce the results of Fig.4 of the main text, we adjusted the amplitudes while those of the left two holes of the same size were set to the same values.

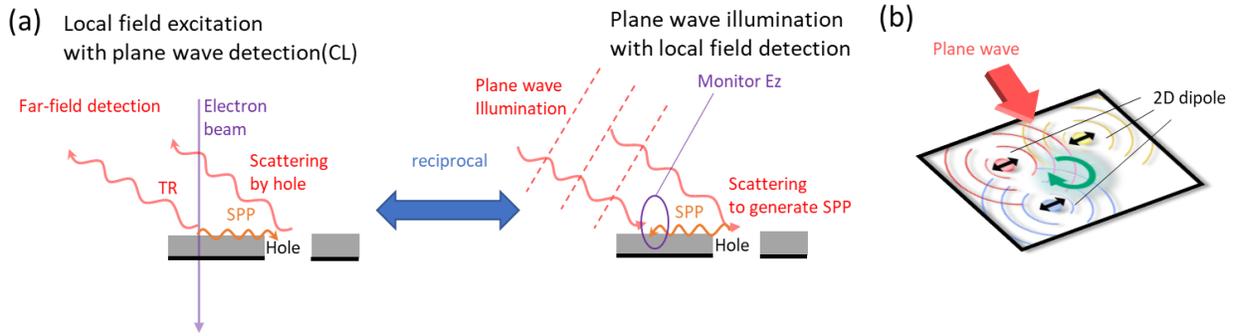

**Figure S2.** (a) Illustration of the reciprocity between far-field cathodoluminescence detection (left) and field-monitoring with plane wave illumination (right). (b) Schematics of the calculation model.

## S3. Calculated phase rotation at singularity position

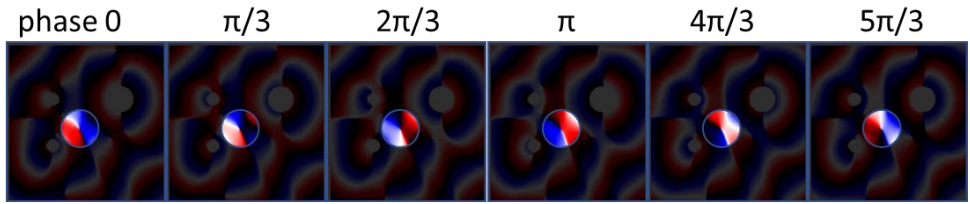

**Figure S3.** Calculated phase distribution plot of the *s*-polarized field at different off set phase values to see the phase rotation around the singularity position. To focus on this spot, other areas are shaded.

## S4. CL spectrum of a hole

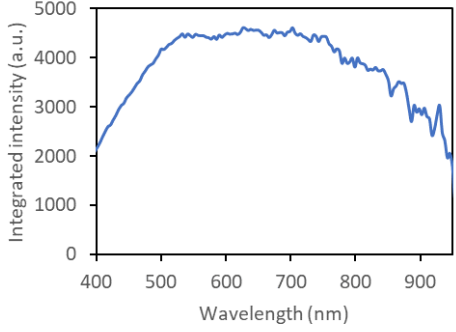

**Figure S4.** CL spectrum of an isolated 500 nm hole integrated over all the detection angles measured with *s*-polarization to eliminate the effect of TR.

## S5. Stokes mapping procedures

The Stokes parameters ($S_{0-3}$) can be obtained from the intensities of the polarized signal $I_\zeta$ (polarization condition: $\zeta$ = s(90°), p(0°), 45°, -45°, RCP, LCP ) as,

$S_0 = I_{\text{non}} = I_s + I_p,$

$S_1 = I_s - I_p = I_{\text{non}}\, p \cos 2\alpha \cos 2\eta,$

$S_2 = I_{45°} - I_{-45°} = I_{\text{non}} p \sin 2\alpha \cos 2\eta = 2 I_s I_p \cos \delta,$

$S_3 = I_{\text{RCP}} - I_{\text{LCP}} = I_{\text{non}}\, p \sin 2\eta = 2 I_s I_p \sin \delta ,$

where $p = \sqrt{S_1^2 + S_2^2 + S_3^2}/S_0$. We define the ellipticity angle $\eta$ and orientation angle $\alpha$ as in Figure 1e. The ellipticity angle $\eta$ ranges from -π/4 to π/4 and is defined such that positive and negative values correspond to RCP and LCP, respectively. The phase difference $\delta$ between the p- and s-polarization fields ($I_{0°}$ and $I_{90°}$) can be calculated as

$\delta = \arg(S_3/S_2).$

We correct the position shift of the image of each polarization, which is caused by sample drifting, and normalize the intensity of the photon map of each polarization. The position of each photon map is adjusted so that the simultaneously obtained STEM images completely overlap. The intensity normalization of each image is also performed for the Stokes calculation, assuming the following two conditions:

1) The integrated intensity over all the mapping area is equal to that of the geometrically symmetric measurement with respect to the *xz*-plane, namely $\iint I_{45°} ds = \iint I_{-45°} ds$ and $\iint I_{RCP} ds = \iint I_{LCP} ds$.

2) The sum of the integrated intensity of orthogonal polarizations should be equal to the non-polarized intensity, i.e. $\iint I_s ds + \iint I_p ds = \iint I_{45°} ds + \iint I_{-45°} ds = \iint I_{RCP} ds + \iint I_{LCP} ds$. These conditions originate from the symmetry of the measurement system with respect to the *xz*-plane. The above formulations deduce simpler normalization relations of the integrated intensities for ±45° and circular polarizations as $\iint I_{45°} ds = \iint I_{-45°} ds = \iint I_{RCP} ds = \iint I_{LCP} ds$.

## S6. Signal intensity

Figure S5a shows the emission probability of TR and SPP on a silver surface calculated for 80 keV electron according to the literature.[1] With the electron beam current of 1 nA, as used in this study, the generated photons by TR are ~ $2 \times 10^6$ /s for the shown wavelength range (400-750nm). The rate of SPPs is calculated as ~ $6 \times 10^6$ /s in the same wavelength range. The number of the detected photons are reduced by selecting the wavelength range and the detection solid angle. As SPPs propagate from the electron beam position toward the hole in a circular manner, the SPP intensity is reduced by the distance as well as by the material loss, as shown in Fig.S5b. For the longer wavelength range (>500 nm), the propagation length is large enough, indicating that the material loss is not significant for this measurement. In contrast, the loss effect is more clearly visible for the plots of 400 nm wavelength in Figs. 2 and 3 in the main text.

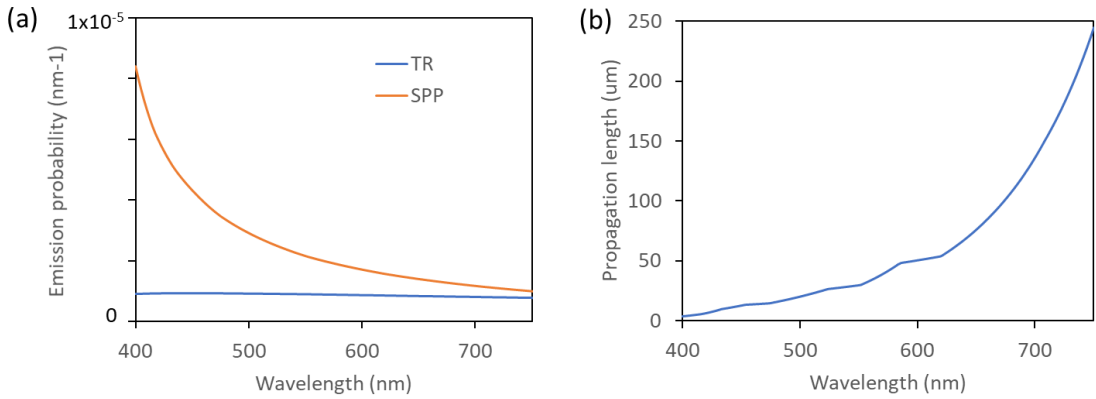

**Figure S5.** (a) Emission probability of TR and SPP on a silver surface calculated for 80 keV electron incidence. (b) Propagation length of the SPP on silver. Literature values for the material parameters are used.[2]